\documentclass{iopart}

\usepackage{iopams}         % Extra math symbols
\usepackage{graphicx}       % Graphics

\begin{document}

    \title[Pion condensation in a two-flavour NJL model]
          {Pion condensation in a two-flavour NJL model:\\the role of charge neutrality}

    \author{J O Andersen and L T Kyllingstad}
    \address{Department of Physics,
        Norwegian University of Science and Technology,
        N-7491 Trondheim,
        Norway}
    \ead{\mailto{andersen@tf.phys.ntnu.no}, \mailto{lars.kyllingstad@ntnu.no}}

    \begin{abstract}
        We study pion condensation and the phase structure in a two-flavour
        Nambu--Jona-Lasinio model in the presence of baryon chemical potential
        $\mu$ and isospin chemical potential $\mu_I$ at zero and finite
        temperature.
        There is a competition between the chiral condensate and
        a Bose-Einstein condensate of charged pions.
        In the chiral limit, the chiral condensate vanishes for any finite
        value of the isospin chemical potential, while there is a charged
        pion condensate that depends on the chemical potentials and the
        temperature.
        At the physical point, the chiral condensate is always nonzero,
        while the charged pion condensate depends on $\mu_I$  and $T$.
        For $T=\mu=0$, the critical isospin chemical potential $\mu_I^c$ for
        the onset of Bose-Einstein condensation is always equal to the pion
        mass.
        For $\mu=0$, we compare our results with chiral perturbation theory,
        sigma-model calculations, and lattice simulations.

%--- Paragraph break added by Lars
        Finally, we examine the effects of imposing electric charge neutrality and
        weak equilibrium on the phase structure of the model.
        In the chiral limit, there is a window of baryon chemical potential
        and temperature where the charged pions condense.
        At the physical point, the charged pions do not condense.
    \end{abstract}


\section{Introduction}
    There has been a tremendous effort in recent years to map out the phase
    diagram of QCD as a function of temperature and baryon chemical 
    potential~\cite{frank,raja,dirk,misha,rob,mark,braunm}.
    It is generally
    accepted that one can calculate the properties of strongly interacting
    matter 
    at asymptotically high temperature or at 
    asymptotically high densities using perturbative QCD.  

    At sufficiently high density and low temperature, we know 
    that QCD is in the colour-flavour locked (CFL) phase. This state is a colour
    superconducting state because the quarks form Cooper pairs in analogy to
    electrons in an ordinary superconductor. 
    In this case, the original symmetry group of QCD,
    $SU(3)_{\rm c}\times SU(3)_L\times SU(3)_R\times U(1)_B$, is broken down
    to $SU(3)_{c+L+R}$ which is a linear combination of the generators of the
    original group. This linear combination locks rotation in colour space to
    rotations in flavour space and this has given the name to the phase.
    In the CFL phase there is an octet of Goldstone modes which arises from 
    the breaking of chiral symmetry and a singlet arising from the breaking
    of the baryon-number conserving group $U(1)_B$. 
    The CFL phase is a superfluid due to the breaking of the global
    $U(1)_B$
    symmetry. This is analogous to the superfluidity encountered in 
    in condensed-matter systems such as $^4$He.
    At very high densities, all nine modes are effectively massless since one can
    ignore the quark masses. This implies that the low-energy properties 
    of the CFL phase can be described in terms of an effective field theory for
    the massless mesons~\cite{mishason,bedaque,kaplan,sjafer}. 
    At moderate densities, one cannot neglect the quark masses
    and chiral symmetry is broken explicitly. 
    Thus only the superfluid mode is exactly massless, while the
    the other mesons acquire masses.
    The lightest massive modes are expected to be the charged and neutral kaons and
    if the chemical potentials are large enough, there is a
    transition to a Bose condensed phase. Bose condensation of kaons in the
    CFL phase has 
    been studied in detail in 
    Refs.~\cite{bub1,forbes,ebertk,ruggers,ebert3,bubver,harmen,alfordus,ebert+,lars2,vietnam}. 

    QCD at finite baryon chemical potential $\mu_B$ is not accessible by
    Monte Carlo simulations due to the complex fermion determinant.
    This is in contrast to QCD at finite isospin chemical potential $\mu_I$
    (still at zero $\mu_B$)  where lattice simulations are possible
    since the functional determinant is real. Thus this is a system
    whose phase diagram one can study on the lattice as a function of 
    a conserved charge. 

    Lattice simulations~\cite{kogut,gupta,misha2}
    suggest that there is a deconfinement transition of pions at high temperature
    and low density, and Bose-Einstein condensation of charged pions at high
    isospin density and low temperature. 
    %$m_{\pi}$~\cite{kogut,gupta,misha2}. 
    In fact, the deconfinement transition and the transition to a charged pion
    condensate seem to coincide. The deconfinement transition is 
    found by measuring the Polyakov loop and the measurements show a sharp
    increase (indicating deconfinement) at approximately the same temperature
    as the onset of pion condensation.

    Pion condensation and the phase diagram of two-flavour QCD have
    been in\-ves\-ti\-ga\-ted using chiral perturbation theory~\cite{son1,split,loewe},
    ladder QCD~\cite{ravag0}, 
    the chiral quark model~\cite{antal},
    the linear sigma models~\cite{kapusta,arthur,china,jens,jenstom,china1},
    NJL models~\cite{china,ravag1,ebert1,ebert2,aussie,china2,china3,abukku},
    and Polyakov-loop NJL models~\cite{muk,ruggi}.
    The PNJL models suggest that deconfinement and the onset of Bose-Einstein
    condensation are two different transitions.
    % --- Lars:
    Very recently, Abuki \textit{et al} \cite{abuki0901} have considered
    the possibility of probing the phase diagram of electrically neutral
    QCD at finite isospin chemical potential using an equilibrated
    gas of neutrinos. They found that a condensate of charged pions arises
    at large enough neutrino densities and small baryon densities, and that
    at even larger neutrino densities there is condensation of charged
    kaons as well.
    % ---

    We note in passing the similarity between three-colour QCD at finite
    $\mu_I$ and two-colour QCD at finite $\mu_B$ \cite{kogut}. 
    The correspondence is
    given by identifying $\mu_I/2$ with $\mu_B$, the charged pion condensate
    with the diquark condensate\footnote{The diquark condensate in 
    two-colour QCD does not break any local symmetries, only global ones.
    The system is therefore a superfluid but not a colour superconductor.}, 
    and the isospin density with the 
    quark density.

    Bose-condensed states or colour-superconducting states may be found in the
    interior of compact stars if the density is high enough.
    In contrast to hadronic matter in heavy-ion collisions, bulk matter
    in compact stars must (on average) be electrically neutral and so a
    neutrality constraint must be imposed~\cite{coulomb,coulomb2}. 
    Similarly, bulk matter must be colour
    neutral and if the system is in a colour superconducting phase, 
    one sometimes has to impose this constraint explicitly.
    It is automatically satisfied if one uses the QCD Lagrangian, but
    this is not so if one describes the system using NJL-type models.
    This is due to the fact that there are no gauge fields in this model and 
    the $SU(N_c)$ colour symmetry is global~\cite{harmen,note,note2,blas}.

    The advantage of models with quarks as microscopic degrees of freedom,
    such as the NJL model, is that one can investigate simultaneously the
    effects of finite baryon chemical potential and isospin chemical potential.
    In the present paper, we consider the two-flavour NJL model at finite
    baryon chemical potential
    and isospin chemical potential. We compute the phase diagram at
    zero and finite temperature as a function of these variables.
    We restrict ourselves to sufficiently low values of the baryon chemical
    potential such that there are no superconducting 
    phases~\cite{coulomb2,huang}.
    We also investigate the 
    effects on the phase diagram by imposing electric charge neutrality 
    and $\beta$-equilibrium. Our work is a generalization of the papers by
    Ebert and Klimenko~\cite{ebert1,ebert2} to finite temperature and finite
    pion mass. 

    The article is organized as follows. In Sec.~\ref{s_lagrangian}, we discuss the 
    Lagrangian and the gap equations of NJL model.
    In Sec.~\ref{phase11}, we discuss the phase diagram at zero as well as finite 
    temperature.
    In Sec.~\ref{s_neutral}, we discuss the issues of charge neutrality and $\beta$-equlibrium.
    In Sec.~\ref{s_nphase}, we investigate the phase diagram 
    at zero and finite temperature imposing charge neutrality
    and $\beta$-equlibrium.
    In Sec.~\ref{discussion}, we summarize and conclude.

\section{Lagrangian and Gap equations} \label{s_lagrangian}
    In this section, we discuss the properties of the Lagrangian of the 
    two-flavour NJL model.
    The Lagrangian can be written as~\cite{buballa}
    \begin{eqnarray}
    {\cal L}&=&{\cal L}_0+{\cal L}_1+{\cal L}_2\;,
    \label{l0}
    \end{eqnarray}
    where the various terms are
    \begin{eqnarray}
        {\cal L}_0 &=&
            \bar\psi(i\gamma^\mu\partial_\mu - m_0)\psi\;, \\
        {\cal L}_1 &=&
            G_1 \left[
                (\bar\psi\psi)^2
                + (\bar{\psi}\boldsymbol{\tau} \psi)^2
                + (\bar\psi i\gamma_5\psi)^2
                + (\bar\psi i\gamma_5 \boldsymbol{\tau} \psi)^2 
            \right]\;, \\
        {\cal L}_2 &=&
            G_2 \left[
                (\bar\psi\psi)^2
                - (\bar{\psi}\boldsymbol{\tau} \psi)^2
                - (\bar\psi i\gamma_5\psi)^2 
                + (\bar\psi i\gamma_5 \boldsymbol{\tau} \psi)^2 
            \right]\;.
    \end{eqnarray}
    Here, $m_0$ is the quark-mass matrix, which is diagonal in flavour space and
    contains the bare quark masses $m_u$ and $m_d$.
    Moreover, ${\boldsymbol \tau}=(\tau_1,\tau_2,\tau_3)$ where 
    $\tau_i$ ($i=1,2,3$) are the Pauli matrices.
    $G_1$ and $G_2$ are coupling constants. 
    The quark field $\psi$ is an isospin doublet
    \begin{eqnarray}
    \psi=
    \left(\begin{array}{c}
    u\\
    d\\
    \end{array}\right)\;.
    \label{d0}
    \end{eqnarray}
    In the following, 
    we take $m_u=m_d$. The Lagrangian~(\ref{l0}) has a global $SU(N_c)$ symmetry
    as well as a $U(1)_B$ baryon symmetry. The latter reflects baryon number conservation.
    In the chiral limit, the Lagrangian~(\ref{l0})
    has an $SU(2)_L\times SU(2)_R$ symmetry.
    Away from the chiral limit, this symmetry is reduced to $SU(2)_{L+R}$
    isospin symmetry.
    ${\cal L}_1$ has an additional $U(1)_A$ axial symmetry. ${\cal L}_2$
    is 't Hooft's instanton-induced interaction term and breaks explicitly
    the $U(1)_A$ axial symmetry of ${\cal L}_1$~\cite{tuft,kobama}. 
    In the following, we shall
    limit ourselves to study the standard NJL Lagrangian by choosing
    $G_1=G_2\equiv G/2$~\cite{buballa} and so Eq.~(\ref{l0})
    reduces to 
    \begin{eqnarray}
    {\cal L} &=& \bar\psi(i\gamma^\mu\partial_\mu - m_0)\psi
    +G\left[ (\bar\psi\psi)^2 
    +(\bar\psi i\gamma_5 \boldsymbol{\tau} 
    \psi)^2 
    \right]\;.
    \label{lag}
    \end{eqnarray}
    We can characterize the system described by the Lagrangian~(\ref{lag})
    by the expectation values of the different conserved charges associated
    with the continuous symmetries. For each conserved charge
    $Q_i$, we introduce a chemical potential $\mu_i$. Note, however, that it is 
    possible to specify the expectation values of different charges simultaneously
    only if they commute. In the present case, we introduce a chemical potential
    $\mu_B$ associated with the $U(1)_B$ baryon symmetry, as well as a chemical
    potential $\mu_I$ associated with the third component of the 
    $SU(2)_{L+R}$ isospin
    group. This is done by adding to the Lagrangian~(\ref{lag}), the terms
    \begin{eqnarray}
    {\cal L}_{B}&=&\mu_B\bar{\psi}\gamma^0B\psi\;,\\ 
    {\cal L}_I&=&
    \mu_I\bar{\psi}\gamma^0I_3\psi\;,
    \end{eqnarray}
    where 
    $B={\rm diag}(1/3,1/3)$
    and $I_3=\tau_3/2$.
    We can then write the Lagrangian as 
    \begin{equation}
    \mathcal{L} =
        \bar{\psi} \left[
            i\gamma^\mu \partial_\mu - m_0
            + \mu\gamma^0 + \delta\mu\gamma^0\tau_3
        \right]\psi
        + G \left[ (\bar{\psi}\psi)^2 + (\bar{\psi} i\gamma^5 \tau_i \psi)^2 \right]\;,
    \label{nlag}
    \end{equation}
    where we have defined the quark chemical potential $\mu$ 
    as well as $\delta\mu$ by
    \begin{eqnarray}
            \mu &\equiv&\frac{\mu_B}{3}\;,\\
    \delta\mu&\equiv&{\mu_I\over2}\;.
    \end{eqnarray}
    In the remainder of the paper we assume $\mu\geq0$ and $\delta\mu\geq0$
    for simplicity.
    In terms of the chemical potentials for the $u$ and the $d$-quarks, $\mu_u$ and $\mu_d$, 
    the chemical potentials $\mu$ and $\delta\mu$ can be written as 
    \begin{eqnarray}                            
    \mu&=&{1\over2}(\mu_u+\mu_d)\;,\\
    \delta\mu&=&{1\over2}(\mu_u-\mu_d)\;.
    \label{dmu}
    \end{eqnarray}
    In the chiral limit, 
    the inclusion of the isospin chemical potential breaks the 
    $SU(2)_L\times SU(2)_R$-symmetry of the Lagrangian to $U(1)_L\times U(1)_R$.
    At the physical point, it breaks the $SU(2)_{L+R}$-symmetry down to
    $U(1)_{L+R}$.

    From the path-integral representation of the free energy density $\Omega$
    \begin{eqnarray}
    e^{-\beta V\Omega}&=&\int{\cal D}\psi^*{\cal D}\psi 
    e^{-\int_0^{\beta}d\tau\int d^3x\cal L}\;,
    \end{eqnarray}
    the expression for the charge density $Q_i$ associated with the
    chemical potential $\mu_i$ can be written as
    \begin{eqnarray}
    Q_i&=&-{\partial{\Omega}\over\partial\mu_i}\,.
    \end{eqnarray}
    We next introduce the auxiliary fields $\sigma$ and $\pi_i$ by
    \begin{eqnarray}
    \label{sigmadef}
    \sigma&=&-2G\bar{\psi}\psi\;,\\
    \pi_i&=&-2G\bar{\psi}i\gamma_5\tau_i\psi\;.
    \label{pidef}
    \end{eqnarray}
    The Lagrangian (\ref{nlag}) can now compactly be written as 
    \begin{eqnarray}
    {\cal L} &=&
        \bar{\psi} \left[
            i\gamma^\mu \partial_\mu - m_0
            + \mu\gamma^0 + \delta\mu\gamma^0\tau_3
            - \sigma - i\gamma^5\pi_a\tau_a
        \right]\psi
        \nonumber \\ &&
        - {1 \over 4G} \left( \sigma^2 + \pi_a\pi_a \right)\;.
    \label{newlag}
    \end{eqnarray}
    The original Lagrangian (\ref{nlag}) 
    can be recovered by using the equations of motion for
    the auxiliary fields $\sigma$ and $\pi_i$ to eliminate them from Eq.~(\ref{newlag}).
    The Lagrangian is now bilinear in the quark fields and so we can integrate
    them out exactly. We then obtain the following effective action
    for the composite fields $\sigma$ and $\pi_i$
    \begin{eqnarray}
        S_{\rm eff} &=&
            -{1\over2}N_c{\rm Tr\log}\left[
                i\gamma^\mu \partial_\mu - m_0
                + \mu\gamma^0 + \delta\mu\gamma^0\tau_3 -\sigma-i\gamma^5\pi_a\tau_a
            \right]
            \nonumber \\ &&
            - \int d^3x\int_{0}^{\beta} d\tau {1\over4G}\left(\sigma^2+\pi_a\pi_a\right),
    \label{effa}
    \end{eqnarray}
    where ${\rm Tr}$ denotes the trace and 
    is over Dirac indices as well as space-time.

    We next introduce a nonzero expectation value for 
    the fields $\sigma$ and $\pi_1$ to allow for a chiral condensate and 
    a charged pion condensate.
    The fields are then written as 
    \begin{eqnarray}
    \sigma & = &
        -2G\langle\bar{\psi}\psi\rangle+\tilde{\sigma}\;, \\
    \pi_1 & = &
        -2G\langle\bar{\psi}i\gamma^5\tau_1\psi\rangle+\tilde{\pi}_1\;,
    \end{eqnarray}
    where $\tilde{\sigma}$ and $\tilde{\pi}_1$ are quantum fluctuating
    fields. In the mean-field approximation, we neglect the fluctuations of 
    the quantum fields $\tilde{\sigma}$ and $\tilde{\pi}_1$
    in the functional determinant.
    This approximation coincides with the leading order of the $1/N_c$-expansion, 
    where $N_c$ is the number of colours~\footnote{In fact, every power
    of the quantum fluctuating fields that arises from expanding the functional
    determinant gives an additional factor of $1/\sqrt{N_c}$ and so 
    Eq.~(\ref{effa}) is a convenient way of organizing a $1/N_c$-expansion}. 

    For notational simplicity, we introduce the quantities
    $M$ and $\rho$ given by 
    \begin{eqnarray}
    M&\equiv&m_0-2G\langle\bar{\psi}\psi\rangle\;, \\
    \rho&\equiv&
    -2G\langle\bar{\psi}i\gamma^5\tau_1\psi\rangle\;,
    \end{eqnarray}
    where $M$ is the constituent quark mass.
    Note that in the chiral limit, the chiral condensate breaks the 
    $SU(2)_L\times SU(2)_R$ symmetry spontaneously
    down to $SU(2)_{L+R}$ in the usual manner. 
    %A nonvanishing pion condensate breaks
    %parity as well as the global $SU(2)$ symmetry.
    Moreover, we can always use the remaining $U(1)$-symmetry to rotate away
    any nonzero value of $\langle\bar{\psi}\gamma^5\tau_2\psi\rangle$.

    Note that we have introduced a single chiral condensate and not separate
    chiral condensates $\langle\bar{u}u\rangle$  and $\langle\bar{d}d\rangle$ 
    for the $u$ and the $d$ quarks. We have chosen $G_1=G_2$ and in this
    case the effective action~(\ref{effa}) only depends on the sum of these
    condensates. In fact, it is easy to show that the two chiral condensates
    must be equal.
    This is in contrast to the calculations of Barducci {\it et al}~\cite{ravag1},
    where the authors choose $G_1=G/2$ and $G_2=0$.
    In that case the effective action is not symmetric under permutation of the 
    two chiral condensates and they are different.
    Our choice is motivated by the fact that $U(1)_A$ symmetry is broken in
    QCD due to instanton effects~\cite{thooft1976}. In fact, setting
    $G_1 = G_2 \neq 0$ in the present model means that axial symmetry is
    maximally violated. When $G_2 = 0$, on the other hand, the
    Lagrangian~(\ref{l0}) is $U(1)_A$-symmetric.

    Also note that we take the condensates $M$ and $\rho$ to be spacetime
    independent. Though it has been demonstrated that the phase diagram may
    contain phases where the condensates are non-uniform~\cite{dautry1979},
    also within the framework of the two-flavour NJL model~\cite{broniowski1990,
    sadzikowski2000,sadzikowski2003,sadzikowski2006,partyka2008}, taking
    this possibility into account is beyond the scope of this study. Still,
    it would be interesting to see how these non-uniform phases are affected
    by imposing neutrality constraints.

    Using standard techniques to evaluate the trace and using
    $\Omega=-S_{\rm eff}/\beta V$, where $\Omega$ is the 
    thermodynamic potential and $V$ is the volume of the system, we obtain
    \begin{eqnarray}
        \Omega
        &=&
            {(M-m_0)^2+\rho^2\over4G}
            -2N_c\int{d^3p\over(2\pi)^3}
            \bigg\{
                E_{\rho}^-
                +T\ln\left[1+e^{-\beta(E_{\rho}^--\mu)}\right]
        \nonumber \\ &&
                +T\ln\left[1+e^{-\beta(E_{\rho}^-+\mu)}\right]
                + E_{\rho}^+
                + T\ln\left[1+e^{-\beta(E_{\rho}^+-\mu)}\right]
        \nonumber \\ &&
                + T\ln\left[1+e^{-\beta(E_{\rho}^++\mu)}\right]
            \bigg\} \;,
    \label{epot}
    \end{eqnarray}
    where the energy $E^{\pm}_{\rho}$ is defined by
    \begin{eqnarray}
    E^{\pm}_{\rho}&=&\sqrt{(E^{\pm})^2+\rho^2}
    \end{eqnarray}
    where
    \begin{eqnarray}
    E^{\pm}&=&E\pm\delta\mu\;,\\ 
    E&=&\sqrt{p^2+M^2}\;. 
    \end{eqnarray}
    In the limit $T\rightarrow0$, the thermodynamic potential~(\ref{epot}) 
    reduces to
    \begin{eqnarray}
        \Omega &=&
            {(M-m_0)^2+\rho^2\over4G}
            -2N_c\int{d^3p\over(2\pi)^3}
            \bigg\{
                E_{\rho}^-
                + (\mu-E_{\rho}^-)\theta(\mu-E_{\rho}^-)
        \nonumber \\ &&
                + E_{\rho}^+
                + (\mu-E_{\rho}^+)\theta(\mu-E_{\rho}^+)
            \bigg\}\;,
    \label{epot0}
    \end{eqnarray}
    which is in agreement with the result of Ref.~\cite{ebert1}.
    The values of $M$ and $\rho$ are found by minimizing
    the thermodynamic
    potential $\Omega$, that is by solving the following gap equations
    \begin{eqnarray}
    \label{g1}
    {\partial\Omega\over\partial M}&=&0\;,\\ %\hspace{2cm}
    {\partial\Omega\over\partial\rho}&=&0\;.
    \label{g2}
    \end{eqnarray}
    Differentiating the effective potential~(\ref{epot})
    with respect to $M$ and $\rho$, we obtain the gap equations 
    \begin{eqnarray}
        0 &=&
            {M-m_0\over2G}
            - 2 N_c M \int{d^3p\over(2\pi)^3}
            \bigg\{
                {E^+\over EE^+_{\rho}}
                [1 - n(E^+_{\rho}-\mu) - n(E^+_{\rho}+\mu)]
        \nonumber \\ &&
              + {E^-\over EE^-_{\rho}}
                [1 - n(E^-_{\rho}-\mu) - n(E^-_{\rho}+\mu)]
            \bigg\} \;,
        \label{g3} \\
        0 &=&
            {\rho\over2G}
            - 2 N_c \rho \int{d^3p\over(2\pi)^3}
            \bigg\{
                {1\over E^+_{\rho}}
                [1 - n(E^+_{\rho}-\mu) - n(E^+_{\rho}+\mu)]
        \nonumber \\ && 
              + {1\over E^-_{\rho}}
                [1 - n(E^-_{\rho}-\mu) - n(E^-_{\rho}+\mu)]
            \bigg\},
        \label{g4}
    \end{eqnarray}
    where $n$ is the Fermi-Dirac distribution,
    \begin{equation}
        n(E) = \frac{1}{e^{\beta E} + 1}.
    \end{equation}
    Taking the limit $T\rightarrow0$, these equations reduce to those 
    obtained by Ebert and Klimenko~\cite{ebert1}:
    \begin{eqnarray}%\nonumber
    0&=&{M-m_0\over2G}-2N_cM\int{d^3p\over(2\pi)^3}\left\{
    {\theta(E_{\rho}^+-\mu)E^+\over EE_{\rho}^+}
    +{\theta(E_{\rho}^--\mu)E^-\over EE_{\rho}^-}
    \right\}\;,
    \\
    0&=&{\rho\over2G}
    -2N_c\rho\int{d^3p\over(2\pi)^3}\left\{
    {\theta(E_{\rho}^+-\mu)\over E_{\rho}^+}
    +{\theta(E_{\rho}^--\mu)\over E_{\rho}^-}
    \right\}\;.
    \end{eqnarray}

    The dispersion relations for the quasiparticles are determined by the
    zeros of the functional determinant in Eq.~(\ref{effa}). 
    One finds~\cite{ebert1}
    \begin{eqnarray}
    &&E_u=E_{\rho}^--\mu\;,\hspace{1cm}E_d=E_{\rho}^+-\mu \\ 
    &&E_{\bar {u}}=E_{\rho}^-+\mu\;,\hspace{1cm}E_{\bar{d}}=E_{\rho}^++\mu\;.
    \end{eqnarray}
    It is easy to show that the dispersion relations for the $\bar{u}$
    and $\bar{d}$-quarks are always gapped, while the dispersion relations for
    the $u$ and $d$-quarks can be gapped or ungapped depending on the values of
    $\mu$.
    In the chiral limit, 
    it follows directly from the gap equations~(\ref{g3}) and~(\ref{g4}) that
    there are no nonzero values for $\mu$, $\delta\mu$, and $T$
    for which $M$ and $\rho$ are nonzero simultaneously. In the pion-condensed
    phase the dispersion relations for $u$ and $d$-quarks then reduce to
    \begin{eqnarray}
    E_u&=&\sqrt{(p-\delta\mu)^2+\rho^2}-\mu\;\\
    E_d&=&\sqrt{(p+\delta\mu)^2+\rho^2}-\mu\;.
    \label{dmassless}
    \end{eqnarray}
    For the $u$-quark, 
    the dispersion relation is gapped or ungapped according to
    \begin{eqnarray}
    \rho&>&\mu\;,\hspace{1cm}{\rm gapped \,\,\,spectrum}\;,\\
    \rho&=&\mu\;,\hspace{1cm}{\rm ungapped \,\,\,quadratic\,\,\,spectrum}\;,\\
    \rho&<&\mu\;,\hspace{1cm}{\rm ungapped \,\,\,linear\,\,\,spectrum}\;.
    \end{eqnarray}
    The possibility of an ungapped quadratic quark spectrum in the context of
    dense baryonic matter was first discussed in Ref.~\cite{miransky2002}.
    The gaplessness of the $d$-quark is determined by the line that is defined
    implicitly by the equation $\mu=\sqrt{(\delta\mu)^2+\rho^2}$, which follows
    from Eq.~(\ref{dmassless}).
    The window in the pion-condensed phase
    where the fermionic excitations are gapless is coined 
    {\it gapless pion condensation}~\cite{ebert1}.
    The situation here is analogous to what happens in colour 
    superconductivity, except that in this case it is $\delta\mu$ that dictates
    the onset of gaplessness: $\delta\mu=\Delta$, where $\Delta$ is the
    superconducting gap~\cite{mark,wang}.
    Moreover, in colour superconductivity $\delta\mu$ is called a stress parameter
    because it gives rise to a mismatch between the Fermi surfaces of the
    $u$ and the $d$-quarks, which imposes an extra energy cost (stress) on the
    formation of Cooper pairs. As long as the stress parameter is small enough
    compared to $\Delta$, BCS pairing can occur~\cite{wang}.
    In the context of pion condensation, $\mu$
    plays the role as a stress parameter since the mismatch is between the
    Fermi surfaces of the $u$ and $\bar{d}$-quarks.
    Similarly, if the stress parameter $\mu$ is small enough relative to $\rho$,
    pion condensation can occur. We will return to this point below.

    The integrals appearing
    in Eqs.~(\ref{epot}),~(\ref{g3}), and~(\ref{g4}) 
    are ultraviolet divergent and one may impose a three-dimensional
    UV cutoff $\Lambda$ to regulate them. Alternatively, one can introduce a
    %\begin{eqnarray}
    %F(p^2)&=&{\Lambda^2\over\Lambda^2+p^2}\;
    %\end{eqnarray}
    form factor~\cite{alford1},
    which falls off for large momenta.
    This ensures that the integrals converge in the ultraviolet.
    Given an ultraviolet cutoff $\Lambda$, the coupling constant $G$
    and the quark mass $m_0$,
    we can use the Dyson equation for the quark propagator to determine  
    the value for the chiral condensate in the vacuum.
    The Dyson equation reads~\cite{buballa}:
    \begin{eqnarray}
    M=m_0+4N_fN_cG\int{d^3p\over(2\pi)^3}{M\over\sqrt{p^2+M^2}}\;,
    \label{dysoneq}
    \end{eqnarray}
    where $N_f$ is the number of flavours and $N_c$ is the number of colours.
    In the remainder of the paper, we also set $N_f=2$,and $N_c=3$.
    This is simply the gap equation~(\ref{g3}) 
    in the vacuum, i.e. for $\mu=\delta\mu=0$.

\section{Phase diagram} \label{phase11}
    In this section, we study the phase diagram at zero and finite temperature
    with\-out taking charge neutrality into account. Since we are not considering
    colour su\-per\-con\-duc\-ti\-vi\-ty, we restrict ourselves to quark chemical potentials
    $\mu<350$ MeV. Since we are using an ultraviolet cutoff of 
    approximately $650$ MeV 
    (see below), one should not trust
    results for the chemical potentials $\mu$ and $\delta\mu$ above approximately
    $400$ MeV. We therefore do not consider values above $\mu=350$ MeV
    and $\delta\mu=400$ MeV in the calculations.
    The equilibrium values of $M$ and $\rho$ are obtained
    by solving numerically the gap equations~(\ref{g3}) and~(\ref{g4}).

    \subsection{Chiral Limit}
        In the chiral limit, the current quark mass vanishes, 
        $m_0=0$. For the numerical calculations, we
        choose an ultraviolet cutoff $\Lambda=650.9$ MeV and 
        a coupling constant $G=5.04$ (GeV)$^{-2}$. Solving the Dyson 
        equation~(\ref{dysoneq}), this gives a constituent quark mass
        in the vacuum of $M=309.9$ MeV. 

        We mentioned in the previous section that there are no values of $\mu$, $\delta\mu$,
        and $T$ such that $M$ and $\rho$ are simultaneously nonzero.
        In other words, the possible solutions to the gap equations are
        a) $M=\rho=0$, b) $M\neq0,\;\rho=0$ and c) $\rho\neq0,\;M=0$.
        In case a), the full symmetry of the Lagrangian is intact, while in case b)
        the $U(1)_L\times U(1)_R$ symmetry is broken spontaneously
        down to $U(1)_{L+R}$ by the chiral condensate. 
        The breaking of the $U(1)$-symmetry gives rise to a conventional Goldstone 
        mode\footnote{For $\mu=\delta\mu=0$, i.e. in the vacuum, there is a broken
        $SU(2)$-symmetry which gives rise to the three massless pions in the usual manner.}
        In case c) the pion condensate
        breaks parity as well the $U(1)_L\times U(1)_R$ symmetry
        down to $U(1)_{AI_3}$. The latter transforms the left-handed and right-handed
        flavour doublet as $\psi_L\rightarrow e^{i\alpha\tau_3}\psi_L$, 
        $\psi_R\rightarrow e^{-i\alpha\tau_3}\psi_R$. The breaking of the 
        $U(1)$-symmetry gives rise to a conventional
        Goldstone boson with a linear dispersion relation
        for small values of the three-momentum. The ground state is therefore
        a pion superfluid~\footnote{At finite density, Lorentz invariance is broken and
        the number of broken generators need not be the same as the number of 
        Goldstone bosons~\cite{holger,shot,brauner}. Goldstone bosons with quadratic 
        dispersion relations appear and the system is not a superfluid in this
        case.}. The dispersion relations for the composite fields $\sigma$ and
        $\pi$ in Eqs.~(\ref{sigmadef})--(\ref{pidef}) are found by first expanding
        the effective action~(\ref{effa}) to second order in the quantum 
        fluctuating field $\tilde{\sigma}$ and $\tilde{\pi}_i$.
        This gives rise to a $4\times 4$ fluctuation matrix $\Gamma(\omega,p)$, 
        where the
        solutions to $\det\Gamma(\omega,p)=0$ 
        determine the dispersion relations $\omega(p)$.
        Details can be found in Ref.~\cite{ebert2}.
        At finite chemical potential there is a mixing between the fields 
        $\tilde{\pi}_1$ and $\tilde{\pi}_2$
        and one of the linear combinations is massless~\cite{ebert1}.

\begin{figure}[htb]
            \center
            \includegraphics[width=7.8cm]{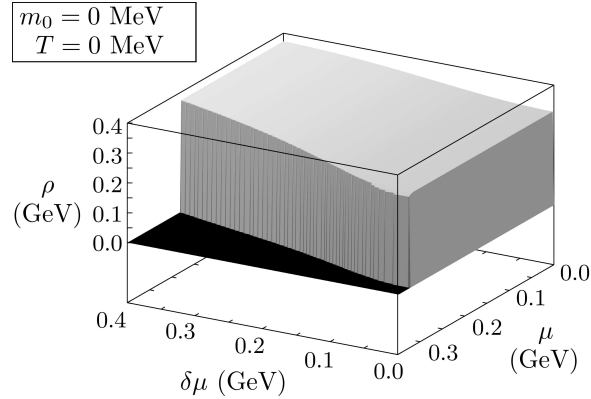}
            \caption{
                Pion condensate in the chiral limit as a function of
                quark chemical potential $\mu$ and $\delta\mu$ at zero temperature.
            }
            \label{t0chiral}
        \end{figure}

        The pion condensate of the two-flavour NJL model at zero temperature is shown in 
        Fig.~\ref{t0chiral}. It is worth noting that along the $\mu$-axis, i.e
        for $\delta\mu=0$, the effective potential no longer depends on $M$ and
        $\rho$ separately, but rather on the combination $M^2+\rho^2$~\cite{ebert2}.
        The effective potential then has the usual mexican-hat shape with infinitely
        many equivalent vacua, where we can choose anyone we wish.
        The chiral condensate can be rotated into pseudoscalar
        condensates via the axial flavour transformations
        \begin{eqnarray}
        \psi\rightarrow e^{i\theta_a\gamma_5\tau_a}\psi\;,
        \hspace{1cm}a=1,2,3\;.
        \label{axrot}
        \end{eqnarray} 
        This is analogous to what happens in colour superconductivity, where
        diquark condensates are rotated into pseudoscalar diquark ones 
        via rotations similar to those in Eq.~(\ref{axrot})~\cite{bub1}.
        At finite quark mass or finite isospin chemical potential, the system becomes
        unstable against developing a nonzero pion condensate.
        At vanishing isospin chemical potential, parity is conserved in QCD and to
        enforce this we  choose $\rho=0$ along the line $\delta\mu=0$. Consequently,
        the chiral condensate is nonvanishing and chiral symmetry is broken
        along the line $\delta\mu=0$. 
        Fig.~\ref{t0chiral} shows that the charged pions condense for 
        any nonzero value of the isospin chemical potential
        %Since $M=0$ in this region, any nonzero value of $\delta\mu$ also
        %restores chiral symmetry~\cite{china,ebert1}.
        This result is in accordance with that of Ebert and Klimenko~\cite{ebert1}
        and calculations 
        in the linear sigma model at finite isospin chemical 
        potential~\cite{china,jens}. 
        Finally, we notice that the phase transition from the pion condensed phase
        to the chirally symmetric phase is first order.
        Notice that the critical isospin chemical potential $\delta\mu$ is decreasing as a
        function of $\mu$.

\begin{figure}[htb]
            \center
            \includegraphics[width=7.8cm]{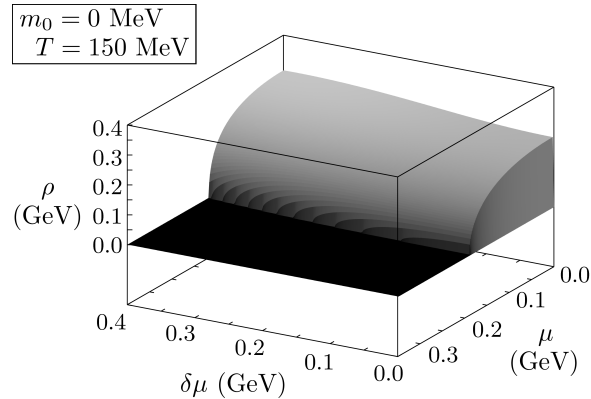}
            \caption{
                Pion condensate in the chiral limit as a function 
                of $\mu$ and $\delta\mu$ for $T=150$ MeV.
            }
            \label{t22chiral}
        \end{figure}

        In Fig.~\ref{t22chiral}, we show the pion condensate as a function of
        quark chemical potential and isospin chemical potential at $T=150$ MeV.
        The region of pion condensation is smaller than at $T=0$ and the 
        transition to a chirally symmetric phase is now second order everywhere.

\begin{figure}[htb]
            \center
            \includegraphics[width=7.8cm]{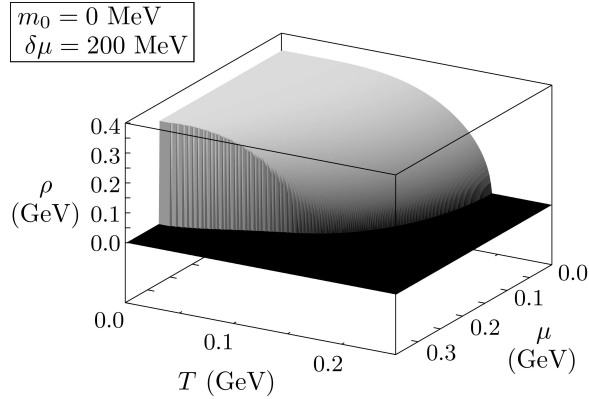}
            \caption{
                Pion condensate in the chiral limit
                as a function quark chemical $\mu$ potential and temperature $T$ for 
                $\delta\mu=200$ MeV.}
            \label{pco0}
        \end{figure}

        In Fig.~\ref{pco0}, we show the
        pion condensate in the chiral limit
        as a function of $\mu$ and $T$ for fixed value of isospin chemical
        potential, $\delta\mu=200$ MeV. As the temperature and the
        quark chemical potential increase, the region
        of pion condensation decreases. 
        For $T=0$, the transition is first order.
        There is a line of first-order transitions starting at $T=0$
        which ends at a critical point given by $T=0.11$ GeV and $\mu=0.22$ GeV.
        The transition is of second order for larger values of $T$
        and smaller values of $\mu$. 

\begin{figure}[htb]
            \center
            \includegraphics[width=7.8cm]{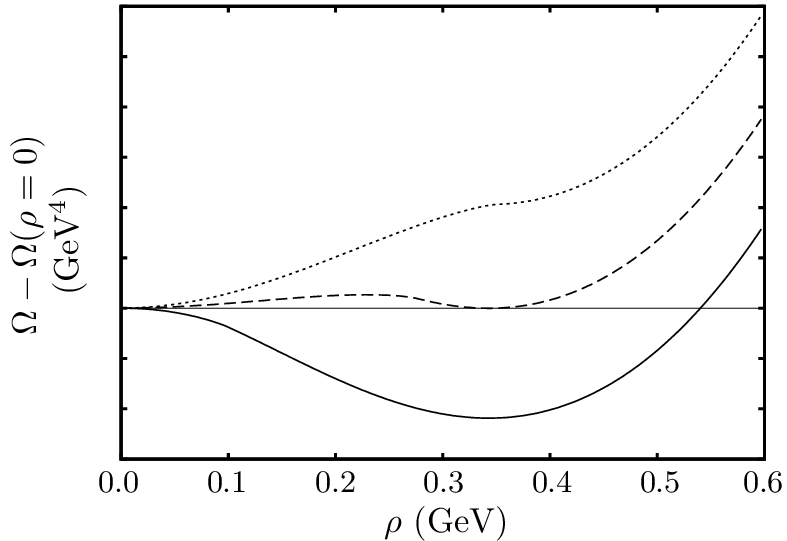}
            \caption{
                $\Omega(\rho)-\Omega(\rho=0)$ in the chiral limit at $T=0$ 
                as a function of $\rho$ for $\delta\mu=200$ MeV and three
                different values of $\mu$.
                Solid line: 100 MeV, dashed line: 275.87MeV, and dotted line: 350 MeV.
            }
            \label{stress}
        \end{figure}

        As mentioned above,
        the quark chemical potential induces a 
        stress in the system and for sufficiently large values of $\mu$ 
        it is no longer energetically favourable to form a Bose condensate of
        charged pions. This is clearly seen in 
        Fig.~\ref{stress}, where we have plotted the thermodynamic potential 
        $\Omega(\rho)$ minus $\Omega(\rho=0)$ for three different values of
        $\mu$ with $\delta\mu=200$ MeV and $T=0$.
        %The solid line is 100 MeV, the dashed line is 275.87, and the 
        %dotted line is 350 MeV. 
        %For values of $\mu$ between 
        %275.87 MeV and approximately 350 MeV, 
        %there is a metastable state with a pion condensate.
        From this figure it is evident that the transition is first order.

    \subsection{Physical point}
        At the physical point, we choose the parameters $m_0=5.5$ MeV, 
        $\Lambda=650.9$ MeV, and $G=5.04\;({\rm GeV})^{-2}$.
        Solving the Dyson equation~(\ref{dysoneq}), this gives 
        a constituent quark vacuum mass $M=325.2$ MeV, and
        the model reproduces the pion mass of $m_{\pi}=140$ MeV.

        Due to the nonzero current quark mass, the chiral condensate $M$ will always be
        nonzero and so chiral symmetry is never restored. For sufficiently high 
        temperature or chemical potentials, the chiral condensate 
        goes towards zero since the
        temperature-independent value of $m_0$ becomes irrelevant.

        The two possible solutions of the gap equations are therefore a) $\rho=0$
        and b) $\rho\neq0$. In the case a) the full symmetry of the Lagrangian 
        is intact, while in case b) parity and 
        the $U(1)_{L+R}$ symmetry are spontaneously broken. In the latter case, there
        is a conventional Goldstone mode and the ground state is a pion superfluid.

        \begin{figure}[htb]
            \center
            \includegraphics[width=7.8cm]{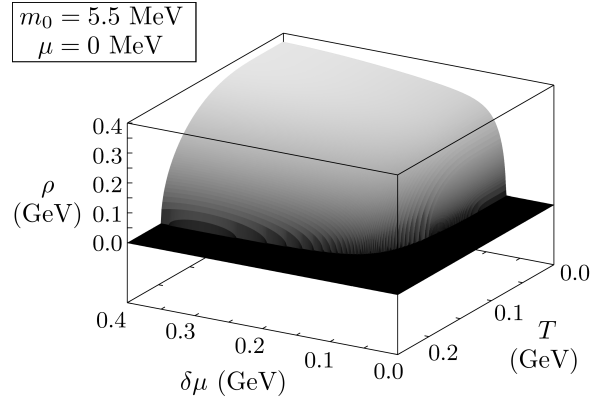}
            \caption{Pion condensate for $\mu=0$ as a function of $T$ and $\delta\mu$.} 
            \label{rhotmu}
        \end{figure}

        In Fig.~\ref{rhotmu}, we show the pion condensate as a function of $T$ and 
        $\delta\mu$ for $\mu=0$. The transition is second order everywhere
        with mean-field critical exponents, which
        is in agreement with the analysis using chiral perturbation 
        theory~\cite{son1,split}. However, lattice calculations~\cite{kogut}
        suggest that the transition is first order for $\mu_I$ large enough.
        The line of first-order transitions ends at a tricritical point and the
        line of second-order transitions extends down to $T=0$.
        This discrepancy between mean-field calculations and lattice calculations
        is most likely due to shortcomings of mean-field theory itself.
        Lattice simulations also suggest that the transition to a Bose-Einstein
        condensed state coincides with the deconfinement phase transition to 
        a quark-gluon plasma.

\begin{figure}[htb]
            \center
            \includegraphics[width=7.8cm]{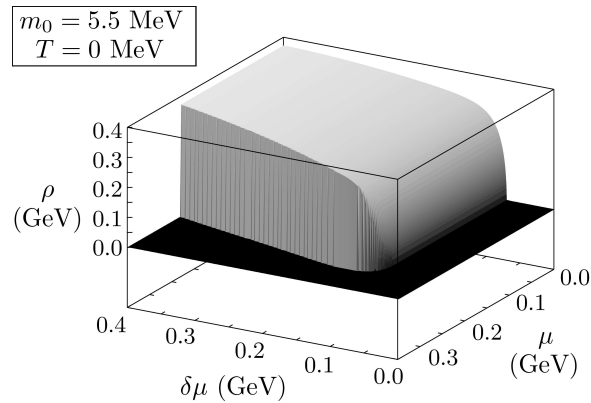}
            \caption{
                Pion condensate at the physical point as a function of quark
                chemical potential $\mu$ and $\delta\mu$ at zero temperature.}
            \label{pco1}
        \end{figure}

\begin{figure}[htb]
            \center
            \includegraphics[width=7.8cm]{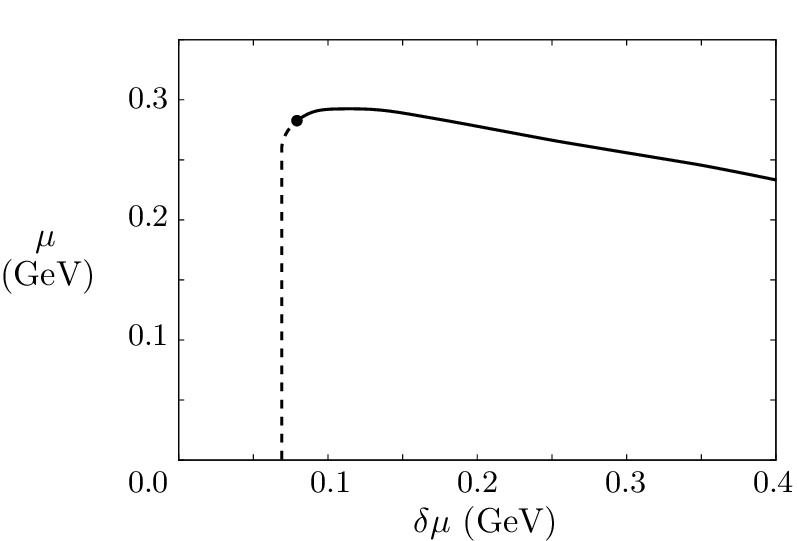}
            \caption{
                Phase diagram at the physical point at $T=0$. The solid curve indicates
                a first-order transition which ends at a critical point.
                The dashed curve indicates a second-order transition.}
            \label{lastfig}
        \end{figure}

        In Fig.~\ref{pco1}, we show the pion condensate as
        a function of $\mu$ and $\delta\mu$ at the physical point for $T=0$.
        Along the axis $\mu=0$, the transition to the pion condensed
        phase occurs at $\delta\mu_{\rm c} = m_\pi/2 = 70$ MeV.
        The phase transition for $\mu=0$ is second order.
        The transition remains second order for $\delta\mu^c$ smaller than 
        approximately $80$ MeV. For larger values of $\delta\mu$ it turns into
        a first-order transition. This is also shown in Fig.~\ref{lastfig}, where
        the solid curve shows a first-order transition
        ending at a critical point, while the dashed line indicates a second 
        order transition. This figure is very similar
        to Fig.~3 of Ref.~\cite{abukku}.

        The results for the pion condensate are in qualitative
        agreement with that obtained by Barducci {\it et al}~\cite{ravag1}.
        Quantitative difference are due to different quark-antiquark interaction
        terms in the Lagrangian as well as different ways of regulating the loop
        integrals.

\begin{figure}[htb]
            \center
            \includegraphics[width=7.8cm]{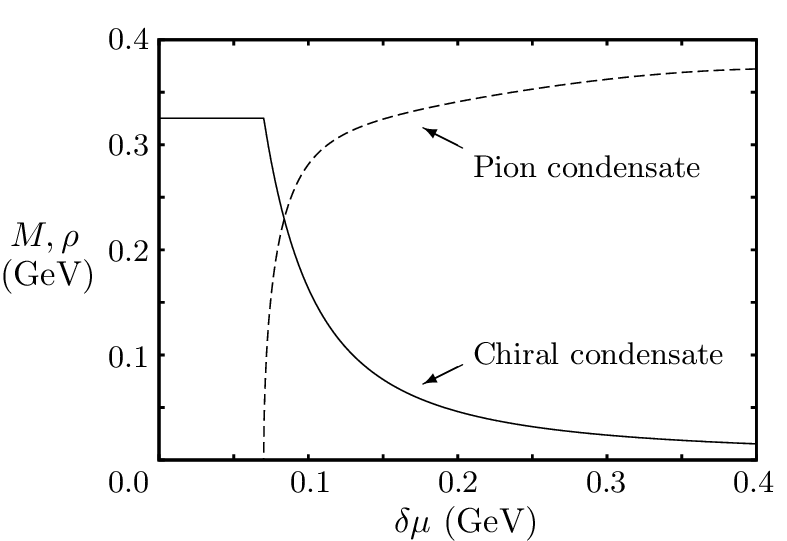}
            \caption{
                Pion condensate (dashed curve) and chiral condensate (solid curve)
                of the two-flavour NJL model at the physical point as a
                function of $\delta\mu$ at $\mu=T=0$.}
            \label{pco3}
        \end{figure}

        In Fig.~\ref{pco3}, we show the chiral and pion condensates as functions of
        the isospin chemical potential for $\mu=T=0$. The chiral condensate
        is constant from $\delta\mu=0$ to $\delta\mu\approx m_{\pi}/2$, after
        which it drops rapidly towards zero at large $\delta\mu$.
        This behaviour is in agreement with the 
        lattice simulations of Kogut
        and Sinclair~\cite{kogut} and the sigma-model 
        calculations of He {\it et al}~\cite{china}.
        We notice the onset of
        pion condensation at $\delta\mu\approx m_{\pi}/2$ and that the condensate
        increases rapidly thereafter. This illustrates the competition
        between the two condensates.

        \begin{figure}[htb]
            \center
            \includegraphics[width=7.8cm]{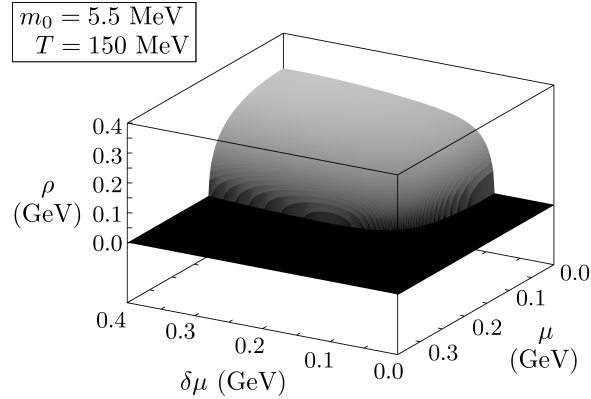}
            \caption{
                Pion condensate of the two-flavour NJL model at the physical point as a
                function of $\mu$ and $\delta\mu$ for $T=150$MeV.}
            \label{pco2}
        \end{figure}

        In Fig.~\ref{pco2}, we show the pion condensate as a function 
        of $\mu$ and $\delta\mu$ at the physical point for $T=150$ MeV.
        The region of Bose condensation becomes smaller as the temperature increases,
        as expected. The phase transition is now second order everywhere.

\begin{figure}[htb]
            \center
            \includegraphics[width=7.8cm]{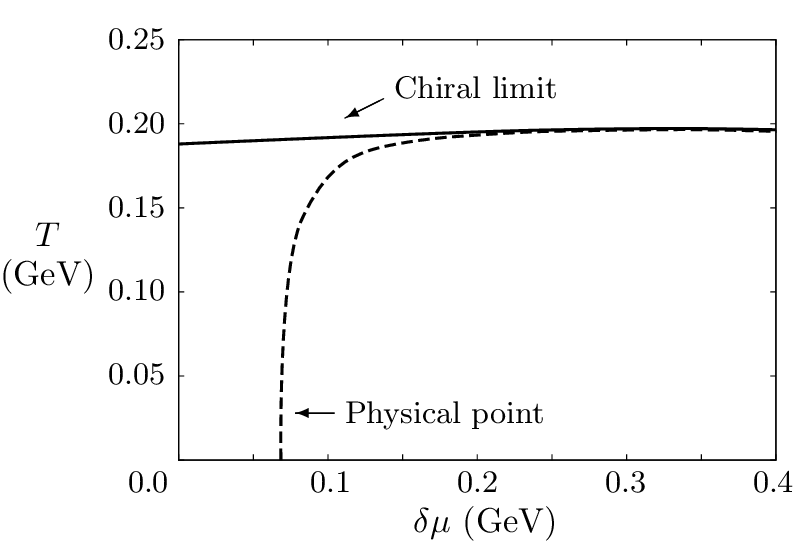}
            \caption{
                Phase diagram as a function of $\delta\mu$ and $T$ for $\mu=0$.
                The solid line is the chiral limit and the 
                dashed line is at the physical point.}
            \label{pco4}
        \end{figure}

        In Fig.~\ref{pco4} we show the phase diagram of the two-flavour NJL model as
        a function of $T$ and $\delta\mu$ for $\mu=0$. The solid line is the chiral
        limit and the dashed line is the at the physical point. We note that the
        two curves are approaching each other for large values of $\delta\mu$, as
        expected. 

\begin{figure}[htb]
            \center
            \includegraphics[width=7.8cm]{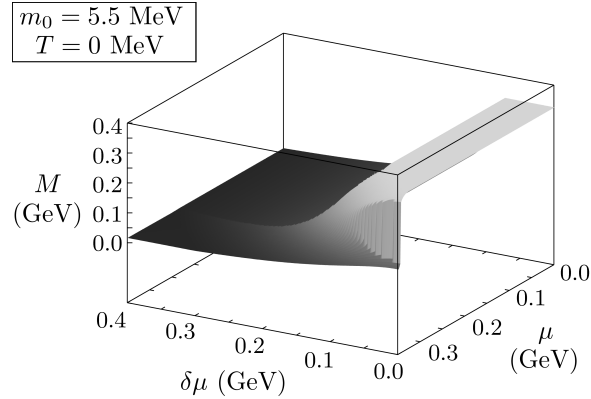}
            \caption{Chiral condensate as a function of 
                quark chemical potential $\mu$ 
                and $\delta \mu$ for zero temperature.}
            \label{chiralcond}
        \end{figure}

        In Fig.~\ref{chiralcond}, we show the 
        chiral condensate as a function of $\mu$ and $\delta\mu$
        at the physical point and for $T=0$. The chiral condensate goes to zero
        as the chemical potentials become large.

\section{Electric Charge Neutrality and $\beta$-equilibrium} \label{s_neutral}
    In the previous section, we have calculated the phase diagram by finding
    the solutions to the gap equations~(\ref{g1}) and~(\ref{g2}). 
    Dense matter inside stars should be neutral with respect to electric
    as well as colour charge, otherwise one would pay an enormous 
    energy price~\cite{coulomb,coulomb2}.
    Since we are not considering colour superconducting phases, colour neutrality
    is automatically satisfied, while we have to impose electric charge 
    neutrality. 

    In addition to charge neutrality, matter should also be in $\beta$ equilibrium,
    that is, weak-interaction processes such as 
    \begin{eqnarray}
    u\leftrightarrow d+e^++\nu\;,
    \label{cheq}
    \end{eqnarray}
    should go with the same rate in both directions. 
    If we assume that the neutrinos can leave the system, their
    chemical potential $\mu_{\nu}$ vanishes. In chemical equlibrium,
    Eq.~(\ref{cheq}) then implies
    \begin{eqnarray}
    \mu_u&=&\mu_d-\mu_e\;.
    \end{eqnarray}
    %where $\mu_u$ and $\mu_d$ are the chemical potentials for the up and down 
    %quarks, respectively. 
    The quark chemical potentials $\mu_u$ and $\mu_d$,  
    and the electron chemical potential $\mu_e$ can be written in terms
    of the quark chemical potential $\mu$ and the electric chemical potential
    $\mu_Q$ as 
    \begin{eqnarray}
    \label{n1}
    \mu_u&=&\mu+{2\over3}\mu_Q\;, \\
    \label{n2}
    \mu_d&=&\mu-{1\over3}\mu_Q\;, \\
    \mu_e&=&-\mu_Q\;,
    \end{eqnarray} 
    and so the system can be described in terms of the
    two independent chemical potentials $\mu$ and $\mu_Q$. 
    In order to impose the constraint of 
    charge neutrality, we additionally require that
    \begin{eqnarray}
    {\partial\Omega\over\partial\mu_Q}&=&0\;.
    \label{nq0}
    \end{eqnarray}
    The constraint~(\ref{nq0}) implies that there is only one independent
    chemical potential, for example $\mu$.
    Once we have picked
    a value for $\mu$, the solutions to the 
    gap equations~(\ref{g1}) and~(\ref{g2}) and the
    neutrality constraint~(\ref{nq0}) determine the chiral condensate $M$, 
    the charged
    pion condensate $\rho$, and the electric chemical potential $\mu_Q$.

    Note that the chemical potential
    appearing in Eq.~(\ref{epot}) is half the
    the sum of the quark chemical potentials
    $\mu_u$ and $\mu_d$, and so according to 
    Eqs.~(\ref{n1}) and~(\ref{n2}), 
    we need to make the substitution 
    $\mu\rightarrow\tilde{\mu}=\mu+\mu_Q/6$. In the remainder, we also 
    replace $\delta\mu$ by $\mu_Q/2$, which follows from
    Eqs.~(\ref{dmu}),~(\ref{n1}), and~(\ref{n2}).

    In the following, we describe the electrons by a noninteracting Fermi gas.
    We then add to the Lagrangian~(\ref{nlag}), the term
    \begin{eqnarray}
    {\cal L}_{\rm electrons}&=&\bar\psi_{e}\left(
    \gamma^{\mu}\partial_{\mu}+\gamma^0\mu_ee-m_e
    \right)\psi_{e}\;,
    \end{eqnarray}
    where $\psi_e$ 
    denotes the electron field, $e$ is the electron charge,
    and $m_e$ is the mass of the electron.
    The thermodynamic potential for the electrons is 
    \begin{eqnarray}\nonumber
    {\Omega}_{\rm electrons}&=&-2\int{d^3p\over(2\pi)^3}
    \left\{E_p+T\ln\left[1+e^{-\beta(E_p-\mu_Q)}\right]
    \right.\\ &&\left.+
    T\ln\left[1+e^{-\beta(E_p+\mu_Q)}\right]
    \right\}\;,
    \label{ev}
    \end{eqnarray}
    where $E_p=\sqrt{p^2+m_e^2}$.
    In the case of massless electrons, one can evaluate the integrals in
    Eq.~(\ref{ev}) exactly and one finds:
    \begin{eqnarray}
    {\Omega}_{\rm electrons}
    &=&-{\mu_Q^4\over12\pi^2}-{\mu^2_QT^2\over6}-{7\pi^2\over180}T^4\;.
    \label{eq}
    \end{eqnarray}
    In the remainder, we neglect the electron mass.
    Adding the electron contribution Eq~(\ref{eq}) to Eq.~(\ref{epot})
    and differentiating with respect to $\mu_Q$, we obtain
    \begin{eqnarray}
        0
        &=&
            {\mu_Q^3\over3\pi^2}+{1\over3}\mu_QT^2
            + \int{d^3p\over(2\pi)^3} \bigg\{
                3 {E^+\over E_{\rho}^+}
                - 3 {E^-\over E_{\rho}^-}
        \nonumber \\ &&
                + \left(1-3{E^+\over E^+_{\rho}}\right)
                    {1\over e^{\beta(E_{\rho}^+ - \tilde{\mu})}+1}
                - \left(1+3{E^+\over E^+_{\rho}}\right)
                    {1\over e^{\beta(E_{\rho}^+ + \tilde{\mu})}+1}
        \nonumber \\ && 
                + \left(1+3{E^-\over E^-_{\rho}}\right)
                    {1\over e^{\beta(E_{\rho}^- - \tilde{\mu})}+1}
                - \left(1-3{E^-\over E^-_{\rho}}\right)
                    {1\over e^{\beta(E_{\rho}^- + \tilde{\mu})}+1}
            \bigg\}\;.
    \label{ch}
    \end{eqnarray}
    In the zero-temperature limit, Eq.~(\ref{ch}) reduces to that 
    obtained by Ebert and Klimenko~\cite{ebert2}:
    \begin{eqnarray}
        0
        &=&
            {\mu_Q^3\over3\pi^2}
            + \int{d^3p\over(2\pi)^3} \bigg\{
                \theta(\tilde{\mu}-E_{\rho}^+)
                +\theta(\tilde{\mu}-E_{\rho}^-)
        \nonumber \\ &&
                + 3\theta(E_{\rho}^+-\tilde{\mu}){E^+\over E^+_{\rho}}
                - 3\theta(E_{\rho}^--\tilde{\mu}){E^-\over E^-_{\rho}}
            \bigg\}
    \;.
    \end{eqnarray}

\section{Phase diagram Revisited} \label{s_nphase}
    In this section, we calculate the phase diagram of the two-flavour NJL model
    as a function of $\mu$ and $T$ imposing the electric charge neutrality
    constraint~(\ref{nq0}). This equation and 
    the gap equations~(\ref{g3}) and~(\ref{g4}) are then solved simultaneously
    to obtain the equilibrium values of $M$ and $\rho$ for the neutral system.
    We are using the same parameter values as in the previous section.

    \subsection{Chiral Limit}
        Again the 
        solutions to the gap equations are
        a) $M=\rho=0$, b) $M\neq0,\;\rho=0$ and c) $\rho\neq0,\;M=0$.

\begin{figure}[htb]
            \center
            \includegraphics[width=7.8cm]{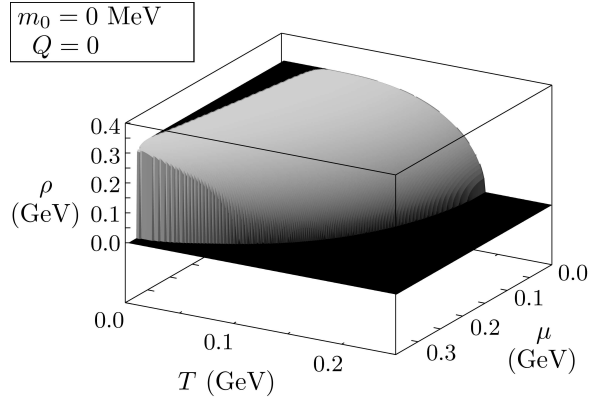}
            \caption{Pion condensate in the chiral limit 
                for neutral matter as a function of 
                quark chemical potential $\mu$ 
                and temperature $T$.}
            \label{tntl}
        \end{figure}

        %In Fig~\ref{tnt00}. 
        %we show the electric chemical potential $\mu_Q$ as a function of the 
        %quark chemical potential $\mu$. 
        %As pointed out in Ref.~\cite{ebert2}, the thermodynamic potential
        %depends on the single variable $\sqrt{M^2+\rho^2}$ if $\mu_Q=0$ and not
        %on $M$ and $\rho$ separately. The potential then has a mexican-hat shape in the
        %$(M,\rho)$-space. One therefore has the freedom to choose the values
        %of $M$ and $\rho$. Since parity is conserved in QCD at $\mu_Q=0$, we 
        %choose $\rho=0$ to ensure this~\cite{ebert2}.
        %\begin{figure}[htb][htb]
        %\includegraphics[width=7.8cm]{n_muq_T0.eps}
        %\includegraphics[width=7.8cm]{img_muQ_chir_neutral}
        %\scalebox{0.5}{\includegraphics{a.eps}}
        %\vspace{-9cm}&
        %\caption{Electric chemical potential $\mu_Q$ as a function of 
        %quark chemical $\mu$ in neutral matter at zero temperature.}
        %\label{tnt00}
        %\end{figure}
        In Fig.~\ref{tntl}, we show the
        pion condensate as a function of quark chemical 
        potential $\mu$ and temperature $T$ for neutral matter.
        The results at $T=0$ agree with those of Ebert and Klimenko~\cite{ebert2}.
        Notice the black wedge starting in the corner 
        $\mu=T=0$. 
        In this area the electric chemical potential $\mu_Q$
        vanishes, giving rise to a nonzero chiral condensate\footnote{
            Again the presence of the chiral condensate in the region
            where $\mu_Q=0$ is due to the fact that the effective
            potential depends on $M^2+\rho^2$ (mexican-hat type potential)
            and we choose $M\neq0$ so that parity is unbroken in the vacuum.
        }.
        %In this area there is nonzero chiral condensate. 
        For $T=0$, the chiral 
        condensate vanishes for quark chemical potentials larger that the
        a critical value of $\mu_{1c}=297$ MeV. 
        %For larger values of the quark chemical potential
        %is nonzero and so chiral symmetry is broken. 
        For a quark chemical potential
        satisfying $\mu_{1c}<\mu<\mu_{2c}$, where $\mu_{2c}=329$ MeV, the
        pion condensate is nonvanishing.
        For chemical
        potentials larger than $\mu_{2c}$, the system is in the normal phase, where
        both condensates vanish. The robustness of this result is discussed in 
        Sec.~\ref{discussion}.
        Finally, we we notice that the transition from the pion-condensed phase to
        the symmetric phase is first order for $T=0$. This first-order line
        ends at a critical point and the transition is second order all the way to
        $\mu=0$.

    \subsection{Physical point}
        Again the two possible solutions of the gap equations are 1) $\rho=0$
        and 2) $\rho\neq0$. It turns out that the only solution is $\rho=0$, i.e.
        there is no charged pion condensate at the physical point.
        In other words, the isospin chemical potential $\mu_I=\mu_Q$ is 
        always smaller than the pion mass. 
        %In Fig.~\ref{muqneutral}, we plot the
        %chemical potential $\mu_Q$ as a function of baryon chemical potential
        %$\mu$ and temperature $T$.

        %\begin{figure}[htb][htb]
        %\includegraphics[width=7.8cm]{n_pion.eps}
        %\includegraphics[width=7.8cm]{img_muQ_phys_neutral}
        %\scalebox{0.5}{\includegraphics{a.eps}}
        %\vspace{-9cm}&
        %\caption{The electric chemical potential $\mu_Q$ at the physical point
        %as a function of $T$ and $\mu$.}
        %\label{muqneutral}
        %\end{figure} 

        This is in accordance with the results of a recent study by 
        Abuki {\it et al}~\cite{abukku},
        where they investigated pion condensation in neutral matter
        as a function of the
        pion mass from the chiral limit all the way to the physical value of $139$ MeV.
        They found a tiny window of pion condensation for pion masses below 
        approximately 10 KeV. Thus pion condensation is very sensitive to the
        explicit chiral symmetry breaking due to a finite quark mass $m_0$.

        \begin{figure}[htb]
            \center
            \includegraphics[width=7.8cm]{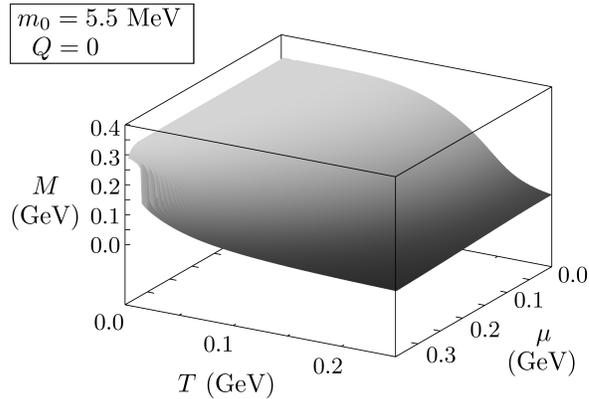}
            \caption{Chiral condensate at the physical point for neutral matter as a function of 
                temperature $T$ and quark chemical potential $\mu$}
            \label{tntl2}
        \end{figure} 

        In Fig.~\ref{tntl2}, we show the chiral condensate for neutral matter
        as a function of temperature and quark chemical potential. The chiral condensate
        decreases with increasing $\mu$ and $T$, but never vanishes.

\section{Summary} \label{discussion}
    In this paper, we have calculated the phase diagram of the two-flavour
    NJL model as a function of the quark and the isospin chemical potential, and
    the temperature in the chiral limit and at the physical point.
    In the chiral limit, we have reproduced the results in 
    Refs.~\cite{ebert1,ebert2} at $T=0$ and generalized them to finite
    temperature.
    At the physical point, we get similar results for the phase structure
    as those obtained in Ref.~\cite{ravag1}.
    The qualitative differences are due to different interaction terms in the
    Lagrangian and a different way of regulating the loop diagrams in the
    gap equations.

    It is natural to ask what happens at larger values of isospin chemical
    potential. We have extended our calculations of Sec.~\ref{phase11} 
    for $\mu=m_0=0$ up to 
    $\delta\mu\approx900$ MeV. These calculations seem to
    indicate that there is a phase transition from a pion condensed phase to a 
    chirally symmetric phase. For zero temperature, this transition is located at
    $\delta\mu\sim880$ MeV.
    However, this result should not be trusted since the UV cutoff
    is $651$ MeV. In fact, there are reasons to believe that there is no phase
    transition as one increases $\delta\mu$. On general grounds, 
    one can show that there is a nonzero
    condensate of the form $\langle\bar{d}\gamma_5u\rangle$ for larger values
    of $\mu_I$~\cite{misha2}. Thus one expects a 
    BEC-BCS type of crossover
    as the quarks bound in the pions become weakly bound
    due to the fact that the QCD running coupling becomes weaker with increasing
    chemical potential $\mu_I$.

    %---- Lars K.
    Our main result comes from imposing
    %----
    %We have also imposed
    the constraints of electric charge neutrality and weak equilibrium.
    In Ref.~\cite{ebert2}, the authors calculate the phases of neutral
    matter with another set of parameters at zero temperature. Using an
    ultraviolet cutoff of $600$ MeV, a coupling constant $G=6.82$ (GeV)$^{-2}$,
    and a constituent quark mass of $400$ MeV, their numerical analysis shows
    that the phase structure differs from the first set of parameters.
    At $T=0$, there is in this case
    no phase with a pion condensate, but a phase transition
    directly from a phase of broken chiral symmetry to a chirally symmetric
    phase at a critical quark chemical potential of $\mu_c=386.2$ MeV.
    This seems to indicate that the window for pion condensation at $T=0$
    that can be seen in Fig.~\ref{tntl} is not a robust result.

    We have seen that the charge neutrality constraint changes the phase diagram
    since it rules out charged pion condensation at the physical point.
    This is true for pion masses larger than approximately 10 KeV \cite{abukku}.
    However, in the presence of a dense neutrino gas, pion condensation
%    at the physical point again becomes a possibility, as shown in
    again becomes a possibility even with realistic values of $m_\pi$, as shown in
    Ref.\ \cite{abuki0901}. In nature this situation arises in supernova
    explosions and possibly at the early stages of the evolution of a neutron star.
    It does not arise in stable matter such as we have considered here,
    because the weakly interacting neutrinos will have had time to leave the system.
%    In fact, charged pion 
%    condensation is ruled out for pion masses larger than approximately 10 KeV.

    %---- Lars K.
    A natural next step would be to investigate the competition between pion
    and kaon condensation in neutral matter using a three-flavour NJL model at finite
    temperature and finite quark chemical potentials $\mu_u$, $\mu_d$, and $\mu_s$.
    Calculations without the neutrality constraint have already been done by
    Barducci \textit{et al}\ \cite{barducci}.
    %----

\ack
    The authors 
    would like to thank T.~Brauner,
    D.~Boer and L.~E.~Leganger
    for useful discussions and
    suggestions.

\end{document}